\documentclass[showpacs,twocolumn,groupedaddress,prl,aps,%longbibliography
]{revtex4-1}

\usepackage{amsmath}
\usepackage{amssymb}
\usepackage{maybemath}
\usepackage{xcolor}
\usepackage{graphicx}
\usepackage{ulem}
\usepackage{bm}
\usepackage{comment}
\usepackage{braket}
\usepackage{mathrsfs}
\usepackage{tikz}

\usetikzlibrary{decorations.pathmorphing}
\usetikzlibrary{shapes}
\usetikzlibrary{shapes.geometric}

\tikzset{
    magnetic/.style={
        fill,
        shape border rotate=-90,
        isosceles triangle,
        isosceles triangle apex angle=60,
        node distance=1,
        minimum height=.1
    }
}

\tikzset{
    othermagnetic/.style={
        fill,
        shape border rotate=90,
        isosceles triangle,
        isosceles triangle apex angle=60,
        node distance=1,
        minimum height=.1
    }
}

\usepackage{dcolumn}
\newcolumntype{.}{D{.}{.}{-1}}

\usepackage[%
  colorlinks=true,
  urlcolor=blue,
  linkcolor=blue,
  citecolor=blue
]{hyperref}

\makeatletter

\newcommand{\Rmnum}[1]{\expandafter\@slowromancap\romannumeral #1@}
\makeatother

\begin{document}

\newcommand{\addrMPIK}{Max Planck Institute for Nuclear Physics, Saupfercheckweg 1, 69117 Heidelberg}

\title{The $g$ factor of bound electrons as a test for physics beyond the Standard Model}

\author{V. Debierre}
\email{vincent.debierre@mpi-hd.mpg.de}
\author{C.~H. Keitel}
\author{Z. Harman}
\email{zoltan.harman@mpi-hd.mpg.de}
\affiliation{\addrMPIK}

\begin{abstract}
  The use of high-precision measurements of the $g$ factor of few-electron ions and its isotope shifts is put forward as a probe for physics beyond the Standard Model. The contribution of a hypothetical fifth fundamental force to the $g$ factor is calculated for the ground state of H-like, Li-like and B-like ions, and employed to derive bounds on the parameters of that force. The weighted difference and especially the isotope shift of $g$ factors are used in order to increase the experimental sensitivity to the new physics contribution. It is found that, combining measurements from four different isotopes of H-like, Li-like and B-like calcium ions at currently accessible accuracy levels, experimental results compatible with King planarity would constrain the new physics coupling constant more than one order of magnitude further than the best current atomic data.
\end{abstract}

%\pacs{}

\maketitle

The $g$ factors of the free electon and free muon have served as precision tests for quantum electrodynamics (QED), the Standard Model (SM) more broadly, and possible extensions of the SM \cite{FreeGTen,Gabrielse,CODATA14,MuonG,MuonGNewP,MuonG2ppm,MuonG2Final}. In parallel, recent years have seen rapid improvements in the experimental \cite{Qui01,Sturm11,Sturm13,Sturm14,Sturm17} and theoretical \cite{BeierOneLoop,YeroGFactor,EveryoneTwoLoop,UlrichTwoLoop,VladZVP,DelbrueckGArticle,CzarneckiSzafron,CzarneckiLetter,SikoraTwoLoop} determination of the $g$ factor of bound electrons (for reviews, see Refs.~\cite{ShabaevReview,ConfReview}), owing to the accelerating development of bound-state quantum field theory \cite{BeierOneLoop,YeroGFactor,EveryoneTwoLoop,UlrichTwoLoop,VladZVP,DelbrueckGArticle,CzarneckiSzafron,CzarneckiLetter,SikoraTwoLoop,MohrOld,MohrPlunienSoff,Shabaev,ForVertexF}. This allows for stringent tests of QED in strong fields, and, to a lesser extent, other sectors of the SM, and can hence be a route for the discovery of phenomena beyond the SM. In the present work, we demonstrate the relevance of the $g$ factor of bound electrons to the search for new physics (NP). A relatively direct method consists in comparing the best available theoretical and experimental results, the largest difference between which allowed by the uncertainties, provides a bound for the NP contribution to the $g$ factor. With this method, we also use data on weighted differences of $g$ factors of ions in different charge states \cite{WDiffVeryOld,WDiffOld,GFactorAlpha,WDiffZ}. Another method is centered on the isotope shift. Precision spectroscopy of the isotope shifts in optical transition frequencies in ions has recently been used \cite{FifthForce,FlambaumIso,PossibleForces,ProbingIS} to examine hypothetical new fundamental forces. When specific candidates for forces are considered, this examination of the isotope shift provides a more direct route to test NP than do high-precision QED calculations. In this work, we generalize this method to $g$-factor precision spectroscopy. Considering four different isotopes of a given ion and two different electronic states, a specific feature in the yet-to-be-obtained $g$-factor isotope-shift data, known as King nonplanarity, could be, with some care, understood as a potential signature of NP. For this conclusion to be warranted, it would need to be assumed that the hypothetical force considered here dominates over other beyond-SM contributions. Conversely, a lack of King nonplanarity in the data would allow the setting of competitive bounds on NP. In order to obtain bounds on the NP parameters, we first derive the correction to the bound-electron $g$ factor due to a hypothetical fifth force.

\textit{Correction to the $g$ factor due to a massive scalar boson.---}New scalar bosons have been proposed as a solution to the long-standing electroweak hierarchy problem~\cite{EightPage}. These massive scalar bosons would carry a fifth force, resulting in an interaction between neutrons and electrons, by coupling to both particle types in a spin-independent way \cite{FifthForce,FlambaumIso,ProbingIS}. The potential exerted on electrons by this hypothetical force is of the Yukawa type \cite{PossibleForces}:
\begin{equation} \label{eq:Potential}
  V_\phi\left(\mathbf{r}\right)=-\hbar c\,\alpha_{\mathrm{NP}}\left(A-Z\right)\frac{\mathrm{e}^{-\frac{m_\phi c}{\hbar}\left|\mathbf{r}\right|}}{\left|\mathbf{r}\right|},
\end{equation}
where $m_\phi$ is the mass of the boson, $\alpha_{\mathrm{NP}}=y_ey_n/4\pi$ is the new-physics coupling constant, with $y_e$ and $y_n$ the coupling of the massive scalar boson to the electrons and the neutrons, respectively, $\hbar$ and $c$ are Planck's reduced constant and the vacuum velocity of light, and $Z$ and $A$ are the number of protons and the total number of nucleons in the nucleus of the considered ion. %{\color{blue}Recall that $\hbar/c=\alpha a_0 m_e$. Hence, three mass regimes can be identified: (\rmnum{1}) the small-mass regime $m_\phi\ll Z\alpha m_e$ for which the interaction potential is very well approximated by a Coulomb-type potential in the range of the electronic wave functions, and where the influence of the fifth force would only correct the fine-structure constant. (\rmnum{2}) the intermediate mass regime $Z\alpha m_e\ll m_\phi\ll \hbar/r_Nc$ with $r_N$ the nuclear radius, which is the most interesting regime: the interaction probes the detailed structure of the electronic wave function. (\rmnum{3}) the heavy-mass regime $m_\phi\gg \hbar/r_Nc$ for which the interaction potential is well approximated by a Dirac delta potential and the influence of the fifth force becomes indistinguishable from finite-size nuclear corrections. (This part in blue could be removed/shortened as this is explained in Ref.~\cite{FifthForce})}
\begin{figure}[b!]
  \begin{tikzpicture}[ultra thick]
    \draw (-2,-.075) -- (2,-.075);
    \draw (-2,.075) -- (2,.075);

    \fill (-1,0) circle (.1);
    \fill (1,0) circle (.1);

    \draw[decorate,decoration=snake] (-1,.125)  -- (-1,1);
    \draw[dashed] (1,.125)  -- (1,1);

    \draw (-1,1.125) node [magnetic]{};

    \fill (.875,1) -- (.875,1.25) -- (1.125,1.25) -- (1.125,1) -- (.875,1);
  \end{tikzpicture}
  \caption{Feynman diagram corresponding to the hypothetical NP contribution to the $g$ factor of a bound electron. The double line represents the bound electron, the wavy line terminated by a triangle denotes a photon from the external magnetic field, and the dashed line terminated by a square denotes a scalar boson from the nuclear neutrons. \label{fig:FDiagram5}}
\end{figure}
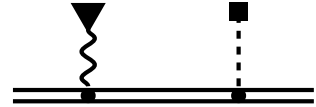

The first-order correction to the $g$ factor of a bound electron in the quantum state $a$ due to this potential is given by the diagram in Fig.~\ref{fig:FDiagram5}, together with the one in which the order of the two interactions is swapped. %Let us write $\ket{\delta a}$ the correction to the bound eigenstate $\ket{a}$ due to an external, static, homogeneous magnetic field $\mathbf{B}=B\mathbf{e}_z$. 
Together they yield
\begin{multline} \label{eq:GCorrStart}
  %=\frac{1}{\mu_0 B m_a}\left(\bra{a}\hat{V}_\phi\ket{\delta a}+\bra{\delta a}\hat{V}_\phi\ket{a}\right)\\
  \Delta g_a=-2\alpha_{\mathrm{NP}}\left(A-Z\right)\frac{\hbar c}{\mu_0 B m_a}\int_0^{+\infty}\mathrm{d}r\,r\,\mathrm{e}^{-\frac{m_\phi c}{\hbar}r}\\
  \times\left[g_a\left(r\right)X_a\left(r\right)\vphantom{\left(\frac{1}{2m_e}\right)}+f_a\left(r\right)Y_a\left(r\right)\right],
\end{multline}
where $\mu_0=e\hbar/2m_e$ is the Bohr magneton, $m_a$ the magnetic projection quantum number of state $a$, $B$ the magnitude of the external, static, homogeneous magnetic field, and $X_a$ and $Y_a$ the corrections to the large ($g_a$) and small ($f_a$) components of the bound electron radial wave function, due to the interaction with the magnetic field, given in Ref.~\cite{ShabaevVirial}.
\begin{comment}
{\color{purple}Eq.~(\ref{eq:GCorrStart}) can be rewritten as
\begin{widetext}
\begin{multline} \label{eq:GCorrEnd}
  \Delta g_a=2\alpha_{\mathrm{NP}}\left(A-Z\right)\lambdabar_e\frac{\kappa_a^2}{j_a\left(j_a+1\right)}\int_0^{+\infty}\mathrm{d}r\,r\,\mathrm{e}^{-\frac{m_\phi c}{\hbar}r}\left[\left(\frac{1}{2}-\kappa_a\right)g_a^2\left(r\right)+\left(\frac{1}{2}+\kappa_a\right)f_a^2\left(r\right)+2\frac{r}{\lambdabar_e}f_a\left(r\right)g_a\left(r\right)\vphantom{\left(\frac{1}{2}\pm\kappa_a\right)}\right],
\end{multline}
\end{widetext}
with $j_a$ and $\kappa_a$ the total and Dirac angular momentum numbers, respectively, and $\lambdabar_e=\hbar/m_e c$ the Compton wavelength of the electron.}
\end{comment}
For the H-like ground state $a=1s$, we obtain
\begin{multline} \label{eq:GdState}
  \Delta g_{1s}=-\frac{4}{3}\alpha_{\mathrm{NP}}\frac{\left(Z\alpha\right)}{\gamma}\left(A-Z\right)\left(1+\frac{m_\phi}{2Z\alpha m_e}\right)^{-2\gamma}\\
  \times\left[3-2\frac{\left(Z\alpha\right)^2}{1+\gamma}-\frac{2\gamma}{1+\frac{m_\phi}{2Z\alpha m_e}}\right].
\end{multline}
The full exact results for the $a=2s$ ground state of Li-like ions, and for the $a=2p_{1/2}$ ground state of B-like ions, are given in the Supplemental Material. However, common asymptotics for all $s$ states can be given for simplicity in the nonrelativistic limit: in the intermediate mass regime $Z\alpha m_e\ll m_\phi\ll \hbar/r_Nc$ ($r_N$ is the nuclear radius), we derive
\begin{subequations} \label{eq:MassLimits}
\begin{multline} \label{eq:IntMassLimit}
  \Delta g_{ns}^{\mathrm{nonrel}}\simeq-\frac{16}{3}\alpha_{\mathrm{NP}}\left(A-Z\right)\frac{Z\alpha}{n^3}\\
  \times\left(Z\alpha\frac{m_e}{m_\phi}\right)^2\left[3-16\,Z\alpha\,\frac{m_e}{m_\phi}\right],
\end{multline}
while in the small mass regime $m_\phi\ll Z\alpha m_e$, we obtain
\begin{equation} \label{eq:SmallMassLimit}
  \Delta g_{ns}^{\mathrm{nonrel}}\simeq-\frac{4}{3}\alpha_{\mathrm{NP}}\left(A-Z\right)\frac{Z\alpha}{n^2}.
\end{equation}
\end{subequations}
With the help of our exact results, we will set bounds on NP in what follows. The free parameters to be constrained are the coupling constant $\alpha_{\mathrm{NP}}$ and the boson mass $m_\phi$.

\begin{figure}[tb]
  \includegraphics[scale=.25]{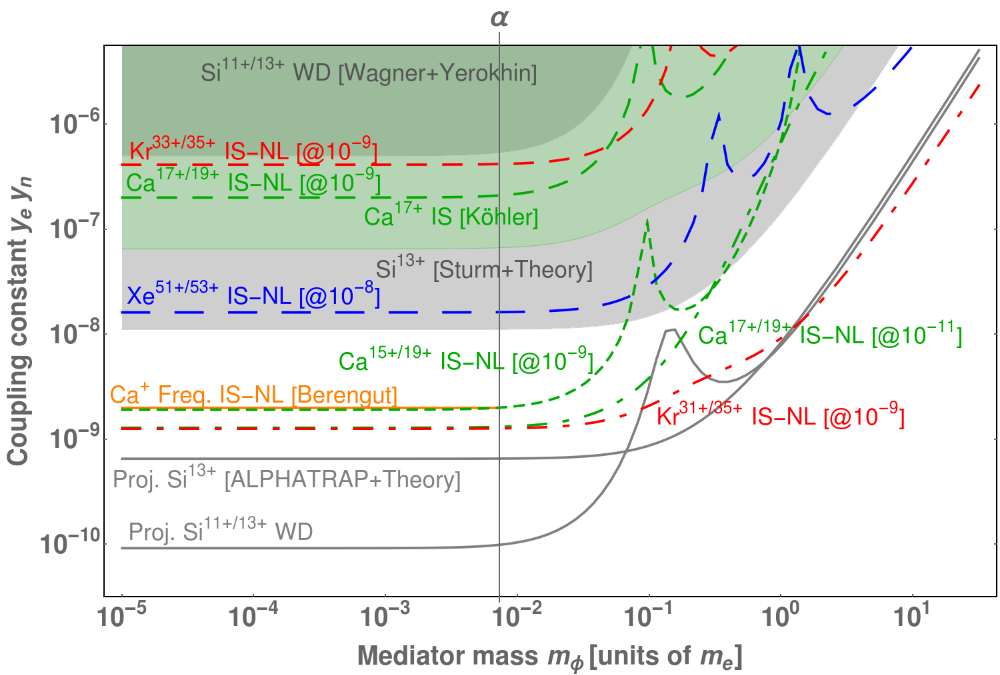}
  \caption{Bounds on the NP coupling constant $y_ey_n=4\pi\alpha_{\mathrm{NP}}$ as a function of the scalar boson mass $m_\phi$. The shaded regions are excluded by available data, and the regions above the curves are excluded by projected data. The thick orange line refers to the small-mass limit of the bound obtained in Ref.~\cite{FifthForce} from isotope shift measurements in transition frequencies in Ca$^+$. Two regions refer to available data on silicon (gray: $^{28}$Si$^{13+}$ ([Sturm+Theory]) \cite{Sturm11}, dark green: $^{28}$Si$^{11/13+}$ ([Wagner+Yerokhin]) \cite{GFactorAlpha,WDiffZ}). The solid gray curves refer to projected bounds for silicon (upper: $^{28}$Si$^{13+}$ \cite{Sturm11}, lower: $^{28}$Si$^{11/13+}$ \cite{GFactorAlpha,WDiffZ}), provided the calculation of radiative corrections is advanced and theory becomes limited by nuclear radius uncertainties. The light green region refers to the calcium isotope shift ($^{40/48}\text{Ca}^{17+}$ ([K{\"o}hler]) \cite{CalciumIS}). All other curves represent projected bounds from a King analysis of the $g$ factor. The green curves refer to projected bounds for calcium [dashed (resp. densely dashed): projections on $^{40/42/44/48}\text{Ca}^{17+/19+}$ assuming King planarity (KP) at the relative experimental accuracy of $10^{-9}$ (resp. $10^{-11}$), dot-dashed: projections on $^{40/42/44/48}\text{Ca}^{15+/19+}$ (KP at $10^{-9}$)]. The red curves refer to krypton [dashed: projections on $^{80/82/84/86}\text{Kr}^{33+/35+}$ (KP at $10^{-9}$), dot-dashed: projections on $^{80/82/84/86}\text{Kr}^{31+/35+}$ (KP at $10^{-9}$)]. The blue curve refers to xenon [dashed: projections on $^{130/132/134/136}\text{Xe}^{51+/53+}$ (KP at $10^{-8}$)].
    \label{fig:Bounds}}
\end{figure}

\textit{Tests with $g$ factors and their weighted difference.---}A first set of bounds can be obtained straightforwardly, by considering that the NP contribution to the $g$ factor is bounded by current uncertainties on the $g$ factor. Let us consider the case of H-like $^{28}\text{Si}^{13+}$. Using Eq.~(\ref{eq:GdState}) and the difference $\Delta g\simeq1.7\times10^{-9}$ \cite{Sturm13,CzarneckiLetter} between theory and experiment (we take the maximum difference allowed by the uncertainties, making use of the updated electron mass uncertainty \cite{Sturm14}), we exclude the larger gray region ([Sturm+Theory]) in Fig.~\ref{fig:Bounds}. The current theoretical uncertainties are $3\times10^{-11}$ from the finite nuclear size correction \cite{ExtractJacek}, and $6\times10^{-10}$ from QED radiative corrections \cite{CzarneckiLetter}, meaning that improvement of the QED theory would be meaningful up to a factor of $20$, yielding the bound represented by the upper solid gray curve (Proj. Si\textsuperscript{13+}) in Fig.~\ref{fig:Bounds} if the agreement with experiments remains at that level of precision.

As this example illustrates, the finite nuclear size correction is a major source of uncertainty in the calculation of bound-electron $g$ factors. To circumvent this, it has been proposed to use weighted differences of $g$ factors of electrons in the $1s$ and $2s$ states \cite{WDiffVeryOld,GFactorAlpha,WDiffZ}, as well as in the $1s$ and $2p_{1/2}$ states \cite{WDiffOld}. These are given \cite{WDiffOld,GFactorAlpha,WDiffZ,ShabaevSize,GlazShabSize} by
\begin{subequations} \label{eq:WDiff}
  \begin{align}
    \delta_{\xi_s}g&=g_{2s}-\xi_s g_{1s},\hspace{12.5pt}\delta_{\xi_p}g=g_{2p_{1/2}}-\xi_p g_{1s},\\
    \xi_s&=2^{-\left(1+2\gamma\right)}\left[1+\frac{3}{16}\left(Z\alpha\right)^2\right],\\
    \xi_p&=2^{-\left(5+2\gamma\right)}\,3\left(Z\alpha\right)^2\left[1+\frac{35}{256}\left(Z\alpha\right)^2\right].
\end{align}
\end{subequations}
In the weighted difference, the finite nuclear size corrections almost entirely cancel each other, through a careful choice of the $\xi$ coefficients. Using the combined experimental and theoretical data on the $1s$--$2s$ weighted difference in $^{28}\mathrm{Si}^{13+}$ from Refs.~\cite{GFactorAlpha,WDiffZ}, we exclude the dark green area ([Wagner+Yerokhin]) in Fig.~\ref{fig:Bounds}. Here, QED theory can be improved \cite{GFactorAlpha,YePaPu} up to a factor of $10^4$ (yielding the bound represented by the lower solid gray curve (Proj. Si$^{11+/13+}$ WD) if the agreement with experiments remains at that level of precision) before the theoretical uncertainty becomes dominated by the finite nuclear size correction. This represents a promising avenue for the setting of more stringent bounds on NP.

\textit{Isotope shifts and the King representation.---}We now turn to the analysis of isotope shifts. Consider two states $1$ and $2$ of a bound electron in an ion. For both states, considering two isotopes $A$ and $A'$ of the same ion, we write \cite{KingBook} the isotope shift in the $g$ factor as
\begin{equation} \label{eq:IsoShift}
g_i^{AA'}=g_i^A-g_i^{A'}\;,\quad i\in\left\{1,2\right\}.
\end{equation}
In the SM, the isotope shift is given at the leading order in the electron-to-nucleus mass ratios $m_e/M_{A^{\left(\prime\right)}}$ and the nuclear radii $\left\langle r^{2}\right\rangle_{A^{\left(\prime\right)}}$ by
\begin{equation} \label{eq:WithinSM}
  g_{i\left(\mathrm{SM}\right)}^{AA'}=K_i\,\mu_{AA'}+F_i\,\delta\left\langle r^{2}\right\rangle_{AA'},
\end{equation}
where $\mu_{AA'}=1/M_A-1/M_{A'}$. The first summand on the r.h.s. is the mass shift, originating from the difference between the masses $M_A$ and $M_{A'}$ of the two isotopes. The second summand is the field shift, and originates from the difference $\delta\left\langle r^{2}\right\rangle_{AA'}=\left\langle r^{2}\right\rangle_A-\left\langle r^{2}\right\rangle_{A'}$ in the nuclear charge radii between the two isotopes. The only quantities that depend on the specific electronic level considered are explicitly referred to as such by the label $i$. Notice that radiative QED corrections to the $g$ factor are largely absent from the isotope shift, as they are approximately the same for all isotopes of a given ion. Taking account of the hypothetical NP contribution (\ref{eq:GCorrStart}) to the $g$ factor, the isotope shift becomes (in analogy to Refs.~\cite{FifthForce,FlambaumIso})
\begin{equation} \label{eq:WithoutSM}
  g_{i\left(\mathrm{NP}\right)}^{AA'}=K_i\,\mu_{AA'}+F_i\,\delta\left\langle r^{2}\right\rangle_{AA'}+\alpha_{\mathrm{NP}}X_i\gamma_{AA'}.
\end{equation}
As can be seen from Eq.~(\ref{eq:GCorrStart}), $\gamma_{AA'}=A-A'$. The parameters of the hypothetical fifth force can be constrained through a study of the isotope shift in the $g$ factor. Indeed, the presence of the third contribution to the isotope shift in Eq.~(\ref{eq:WithoutSM}) can be detected on King plots. Let us introduce $G_i^{AA'}\equiv g_i^{AA'}/\mu_{AA'}$ as well as $h_{AA'}\equiv\gamma_{AA'}/\mu_{AA'}$. The King representation is now constructed from considering four different isotopes $A$, $A'_1$, $A'_2$, $A'_3$. We introduce the notations
\begin{subequations} \label{eq:Vector}
  \begin{align}
    \mathbf{G}_i&\equiv\left(G_i^{AA'_1},G_i^{AA'_2},G_i^{AA'_3}\right),\\
    \mathbf{h}&\equiv\left(h^{AA'_1},h^{AA'_2},h^{AA'_3}\right),\hspace{25pt}\mathbf{A}_{\mathrm{tt}}\equiv\left(1,1,1\right).
  \end{align}
\end{subequations}
\begin{comment}
  {\color{blue}in the SM, we start from Eq.~(\ref{eq:WithinSM}) to derive
\begin{equation} \label{eq:VectorSM}
  \mathbf{G}_{2\left(\mathrm{SM}\right)}=\left(K_2-\frac{F_2}{F_1}K_1\right)\mathbf{A}_{\mathrm{tt}}+\frac{F_2}{F_1}\mathbf{G}_{1\left(\mathrm{SM}\right)}
\end{equation}
while, in the presence of NP, the corresponding relation is
\begin{multline} \label{eq:VectorNP}
  \mathbf{G}_{2\left(\mathrm{NP}\right)}=\left(K_2-\frac{F_2}{F_1}K_1\right)\mathbf{A}_{\mathrm{tt}}\\
  +\frac{F_2}{F_1}\mathbf{G}_{1\left(\mathrm{NP}\right)}+\alpha_{\mathrm{NP}}\mathbf{h}X_1\left(\frac{X_2}{X_1}-\frac{F_2}{F_1}\right).
\end{multline}
This establishes that (all this part in blue might be removed as similar expressions are present in Ref.~\cite{FifthForce})}
\end{comment}
King nonplanarity is measured \cite{FifthForce} by the parameter
\begin{equation} \label{eq:Non}
  \mathcal{N}\equiv\frac{1}{2}\left(\mathbf{G}_1\times\mathbf{G}_2\right)\cdot\mathbf{A}_{\mathrm{tt}}
\end{equation}
and, while, in the SM, it is easily seen that $\mathcal{N}_{\mathrm{SM}}=0$ at the leading order, in the presence of NP, the nonplanarity reads
\begin{equation} \label{eq:NonNP}
  \mathcal{N}_{\mathrm{NP}}=\frac{\alpha_{\mathrm{NP}}}{2}\left(\mathbf{A}_{\mathrm{tt}}\times\mathbf{h}\right)\cdot\left(X_1\mathbf{G}_{2\left(\mathrm{NP}\right)}-X_2\mathbf{G}_{1\left(\mathrm{NP}\right)}\right).
\end{equation}
Hence nonplanarity in the King representation is a possible sign of NP. However, other corrections cause departures from planarity. Indeed the SM contributions (\ref{eq:WithinSM}) to the isotope shift are only valid at the leading order. Subleading (nuclear) SM contributions to the isotope shift are another source of nonplanarity. As such, one should be careful before interpreting potential observed nonplanarities as a sign of NP.

\textit{Standard model contributions to King nonplanarity.---}Several subleading nuclear corrections to the $g$ factor induce King nonplanarity in the SM, which somewhat limits the range of nonplanarity as a signature of NP. These corrections are a higher-order contribution to the finite nuclear size correction \cite{FNSGHigher}, the nuclear polarization correction \cite{NuPolAna,NuPolNum}, the nuclear deformation correction \cite{NuDefJacek,NuDefNiklas}, which all contribute to the field shift, and the higher-order mass correction \cite{HOMass}. We have evaluated these contributions using Eq.~(15) of Ref.~\cite{FNSGHigher}, Eq.~(26) of Ref.~\cite{FlambaumIso} [which provides a simple model for the nuclear polarization correction, and tends to overevaluate the more precisely calculated correction \cite{NuPolAna} by a factor of $\sim2$, making our planarity test quite conservative], Eqs.~(8)--(12) of Ref.~\cite{NuDefJacek}, and Eq.~(47) of Ref.~\cite{HOMass}. We take the nuclear charge radii values found in Ref.~\cite{NuclRadiiUpdate} and the nuclear deformation parameters found in Ref.~\cite{NuclDefValues}. The higher-order finite nuclear size correction is not given in the literature for $p$ states: we built an upper bound for $2p_{1/2}$ by considering that it scales, with respect to the leading-order, in the same way that it does for $2s$, with an extra factor of~10.

\textit{Tests with the isotope shift.---}A first set of bounds on NP can be obtained by considering simple isotope shifts, without recourse to the King formalism. The shift in the $g$ factor between the $^{40}\text{Ca}^{17+}$ and $^{48}\text{Ca}^{17+}$ isotopes of Li-like calcium \cite{CalciumIS,RecoilLithium} has been calculated to be $\Delta g_{\mathrm{th}}=11.056\,\left(16\right)\times10^{-9}$ in the SM, while the experimental value is $\Delta g_{\mathrm{xp}}=11.70\,\left(1.39\right)\times10^{-9}$, meaning that the error bars allow for a maximum difference of $\Delta g\simeq2.05\times10^{-9}$. With the NP correction (\ref{eq:GCorrStart}) to the $g$ factor in the $2s$ state, this leads to the exclusion of the light green region ([K\"ohler]) in Fig.~\ref{fig:Bounds}.
%\begin{figure}[tb]
  %\includegraphics[scale=.25]{2P_Talk.png}
  %\caption{Same as Fig.~\ref{fig:Bounds}. The green curves refer to calcium [dot-dashed (resp. dotted): projections on $^{40/42/44/48}\text{Ca}^{15+/19+}$ assuming KP at the experimental accuracy level of $10^{-9}$ (resp. $10^{-11}$)]. The red curve refers to krypton [dot-dashed: projections on $^{80/82/84/86}\text{Kr}^{31+/35+}$ (KP at $10^{-9}$)].
    %\label{fig:BoundsP}}
%\end{figure}

Bounds on NP from King planarity tests are obtained in the following way: the nonplanarity parameter (\ref{eq:Non}) computed from (in our case, simulated) experimental data is first compared to its first-order propagated error $\sigma_{\mathcal{N}}$, as explained in Ref.~\cite{FifthForce}. If $\mathcal{N}<\sigma_{\mathcal{N}}$, the data is considered planar, and we use the first-order propagated error $\sigma_{\alpha_{\mathrm{NP}}}$ as the upper bound for $\alpha_{\mathrm{NP}}$. To compute $\sigma_{\alpha_{\mathrm{NP}}}$, we have generalized Eq.~(\ref{eq:WithoutSM}), taking account of the subleading nuclear corrections to the $g$ factor that induce King nonplanarity in the SM:
\begin{equation} \label{eq:WithinWithout}
  g_{i\left(\mathrm{TN}\right)}^{AA'}=K_i\,\mu_{AA'}+F_i\,\delta\left\langle r^{2}\right\rangle_{AA'}+\alpha_{\mathrm{NP}}X_i\gamma_{AA'}+s_{i\,AA'}.
\end{equation}
Here the subscript $\left(\mathrm{TN}\right)$ indicates that the total King nonplanarity is captured, including its SM contributions. Note that $s_{i\,AA'}$, the contribution to the isotope shift from these subleading nuclear corrections, cannot be factorized as the product of a nuclear term and an electronic term. Setting $S_{i\,AA'}\equiv s_{i\,AA'}/\mu_{AA'}$ and
\begin{equation} \label{eq:SVector}
    \mathbf{S}_i\equiv\left(S_i^{AA'_1},S_i^{AA'_2},S_i^{AA'_3}\right),
\end{equation}
we obtain
\begin{comment}
\begin{widetext}
\begin{multline} \label{eq:Argh}
  \mathcal{N}_{\mathrm{TN}}=\frac{\alpha_{\mathrm{NP}}}{4}\left(\mathbf{A}_{\mathrm{tt}}\times\mathbf{h}\right)\cdot\left[\left(X_1\frac{F_2}{F_1}-X_2\right)\mathbf{G}_{1\left(\mathrm{TN}\right)}-\left(X_2\frac{F_1}{F_2}-X_1\right)\mathbf{G}_{2\left(\mathrm{TN}\right)}\right]\\-\frac{\mathbf{A}_{\mathrm{tt}}}{2}\cdot\left[\mathbf{G}_{1\left(\mathrm{TN}\right)}\times\left(\frac{F_2}{F_1}\mathbf{S}_1-\mathbf{S}_2\right)-\mathbf{G}_{2\left(\mathrm{TN}\right)}\times\left(\frac{F_1}{F_2}\mathbf{S}_2-\mathbf{S}_1\right)\right]
\end{multline}
\end{widetext}
\end{comment}
%\begin{widetext}
\begin{multline} \label{eq:Argh}
  \mathcal{N}_{\mathrm{TN}}=-\frac{\mathbf{A}_{\mathrm{tt}}}{2}\cdot\left[\mathbf{G}_{1\left(\mathrm{TN}\right)}\times\left(\frac{F_2}{F_1}\mathbf{S}_1-\mathbf{S}_2\right)-\left(1\leftrightarrow2\right)\right]\\
  +\frac{\alpha_{\mathrm{NP}}}{4}\left(\mathbf{A}_{\mathrm{tt}}\times\mathbf{h}\right)\cdot\left[\left(X_1\frac{F_2}{F_1}-X_2\right)\mathbf{G}_{1\left(\mathrm{TN}\right)}-\left(1\leftrightarrow2\right)\right]
\end{multline}
%\end{widetext}
which can be checked to simplify to its equivalent Eq.~(\ref{eq:NonNP}) in the $\mathbf{S}_i\rightarrow\mathbf{0}$ limit, where the subleading nuclear corrections are neglected. It is then straightforward from Eq.~(\ref{eq:Argh}), to compute $\alpha_{\mathrm{NP}}$ and its propagated error $\sigma_{\alpha_{\mathrm{NP}}}$. The leading-order mass and field shifts in Eq.~(\ref{eq:WithinSM}) are computed according to Refs.~\cite{RecoilAllOrders,RecoilLithium,RecoilBoron,ShabaevSize}

We thus derive bounds from possible future experiments on the spinless $A=40,\,42,\,44,\,48$ isotopes of calcium. We first use the $1s$ and $2s$ states (the respective ground state of the H-like and Li-like ions), for the explicit realization of our analysis. Assuming that the experimental data for the H-like and Li-like ground states will be compatible with King planarity, and that the relative experimental uncertainties on the measured $g$ factors will be $10^{-9}$ and $10^{-11}$, respectively, we obtain the projected bounds appearing in dashed and densely dashed green in Fig.~\ref{fig:Bounds}. This assumes that the mass uncertainty of the Ca ions will be reduced by up to two orders of magnitude (in the case of a relative experimental precision of $10^{-11}$), which can be achieved by the newly commissionned PENTATRAP Penning-trap setup \cite{PENTA}. We also consider the $A=80,\,82,\,84,\,86$ isotopes of krypton and the $A=130,\,132,\,134,\,136$ isotopes of xenon. With the same reasoning, and with hypothetical data with uncertainties $10^{-9}$ (Kr) and $10^{-8}$ (Xe), respectively, we obtain the projected bounds appearing in red and blue in Fig.~\ref{fig:Bounds}. For better experimental accuracies, King planarity breaks because of the subleading nuclear (SM) corrections, preventing the setting of bounds on NP in the present formalism.

We turn to deriving bounds from the $1s$ and $2p_{1/2}$ states (the respective ground states of the H-like and B-like ions) of the same four isotopes of calcium and krypton. It is expected \cite{FifthForce} that pairs of electronic levels with more dissimilar wave functions can yield better tests, and indeed, we obtain a more stringent bound on the NP coupling constant with H-like/B-like pairs than with H-like/Li-like pairs, assuming a relative experimental accuracy of $10^{-9}$. This is true for both calcium and krypton, as can be seen on Fig.~\ref{fig:Bounds}.

\begin{figure}[tb]
  \includegraphics[scale=.25]{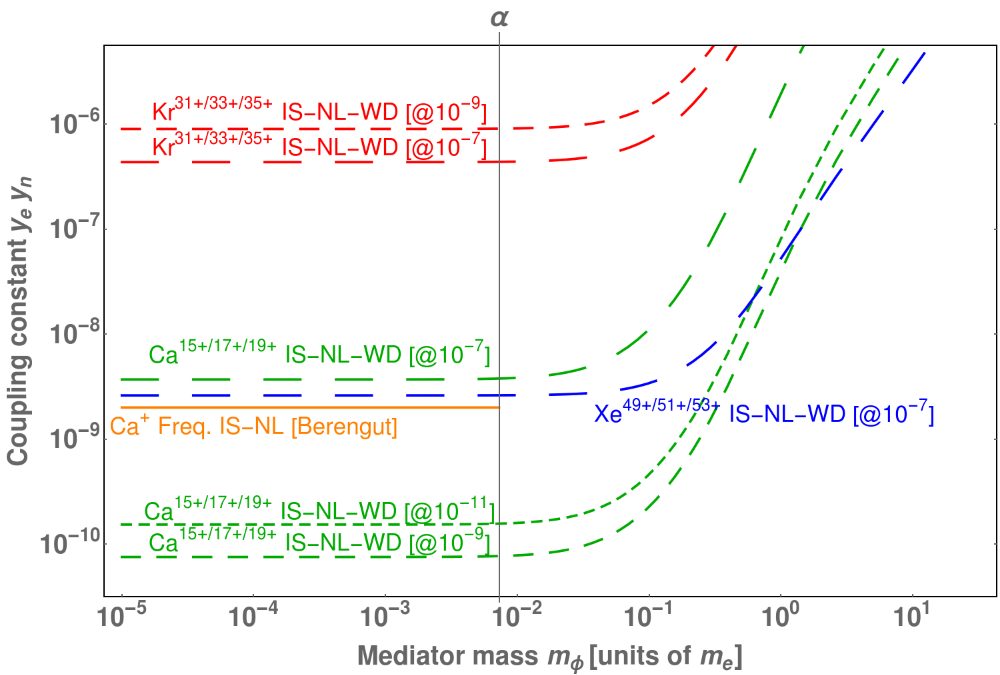}
  \caption{Same as Fig.~\ref{fig:Bounds}, with projected bounds from a King analysis of the weighted differences. The green curves refer to calcium [sparsely dashed (resp. dashed, densely dashed): projections on $^{40/42/44/48}\text{Ca}^{15+/17+/19+}$ assuming KP at the relative experimental accuracy of $10^{-7}$ (resp. $10^{-9}$, $10^{-11}$)], the red curves to krypton [sparsely dashed (resp. dashed): projections on $^{80/82/84/86}\text{Kr}^{31+/33+/35+}$ (KP at $10^{-7}$ (resp. $10^{-9}$))] and the blue curve to xenon [(sparsely dashed: projections on $^{130/132/134/136}\text{Kr}^{49+/51+/53+}$ (KP at $10^{-7}$))].
    \label{fig:WDiff}}
\end{figure}

Finally, the King analysis for isotope shifts can be combined with the weighted difference. We note that, just like the (leading) finite nuclear size correction to the $g$ factor, the higher-order finite nuclear size correction, the nuclear deformation correction and the nuclear polarization correction scale as $1/n^3$ for $ns$ states. The NP correction, however, scales as $1/n^2$ in the low carrier-mass limit [see Eq.~(\ref{eq:SmallMassLimit})], meaning that the $1s$--$2s$ weighted difference will also suppress the subleading nuclear corrections that cause King nonplanarities in the SM, while preserving the NP corrections. This makes the weighted difference a powerful tool to emphasise NP contributions to King nonplanarity. We repeat the same King analysis on the isotope shift with the pair of `modified' $g$ factors defined in Eq.~(\ref{eq:WDiff}). Due to the suppression of the subleading nuclear corrections in the $1s$--$2s$ weighted difference, the obtained bounds are more sensitive to the simulated experimental noise around the values expected from SM calculations. A typical set of bounds, however, is given in Fig.~\ref{fig:WDiff}. It is seen that use of the weighted difference is successful for calcium in particular: with the fairly modest relative experimental accuracy of $10^{-9}$, we find that King planarity would allow the setting of bounds on NP more than one order of magnitude more stringent than that obtained in Ref.~\cite{FifthForce}. All simulated data sets show such an improvement by one to two orders of magnitude.

\textit{Conclusion.---}We have shown that $g$-factor measurements in highly charged ions provide a very competitive framework to obtain bounds on NP. Investigating the influence of a  fifth force, acting between neutrons and electrons, on the $g$ factor of bound electrons, we have used the isotope shift in $g$ factors, as well as the weighted difference technique. We also accounted for subleading nuclear contributions to the isotope shift in the SM. The bounds readily obtained through existing data are between one and two orders of magnitude less stringent than those obtained \cite{ProbingIS} through published measurements of the isotope shift in transition frequencies in Ca$^+$. As we have found, measurement of isotope shifts in the $g$ factor of H-like, Li-like and B-like calcium at the very achievable accuracy of $10^{-9}$, can allow for the setting of bounds more than one order of magnitude more stringent than the ones obtained from the Ca$^+$ frequency shift. Isotope shifts in the $g$ factor of bound electrons can also be used to obtain bounds on other hypothetical interactions \cite{PossibleForces}, such as those arising in models with $B$--$L$ vector bosons \cite{UBLStueck} or in chameleon models \cite{ChameleonSurprise,ChameleonCosmology}.

We thank Klaus Blaum for drawing our attention to the topic of Standard Model tests with atomic experiments, as well as Sven Sturm, Niklas Michel and Natalia~S. Oreshkina for helpful conversations.

\bibliography{../Biblio}
\end{document}